\documentclass[fleqn,graphicx,natbib]{aastex}
\usepackage{spr-astr-addons} 
\usepackage{times}
\usepackage{mathptmx}   
\newcommand\etal{et al.}
\newcommand\cesam{{\sc cesam}}
\newcommand\posc{{\sc posc}}
\begin{document}

\title{Grids of stellar evolution models for asteroseismology (\cesam\ + \posc)}


\author{Jo\~ao P. Marques}
\affil{Centro de Astrof\'{\i}sica da Universidade do Porto,
   Rua das Estrelas, 4150-762 Porto, Portugal \\
   Grupo de Astrof\'{\i}sica da Universidade de Coimbra, Portugal}
   
\author{M\'ario J. P. F. G. Monteiro}
\affil{Centro de Astrof\'{\i}sica da Universidade do Porto,
   Rua das Estrelas, 4150-762 Porto, Portugal \\
   Departamento de Matem\'atica Aplicada da Faculdade de Ci\^encias,
   Universidade do Porto, Portugal}
           
\author{Jo\~ao M. Fernandes}
\affil{Grupo de Astrof\'{\i}sica da Universidade de Coimbra, Portugal \\
   Departamento de Matem\'atica da Faculdade de Ci\^encias e Tecnologia,
   Universidade de Coimbra, Portugal}

\begin{abstract}
In this paper we present a grid of stellar evolution models, computed with an up-to-date physical description of the internal structure, using the {\em Code d'\'Evolution Stellaire Adaptatif et Modulaire} (\cesam).
The evolutionary sequences span from the pre-main sequence to the beginning of the Red Giant Branch and cover an interval of mass typical for low and intermediate mass stars. The chemical composition (both helium and metal 
abundance) is the solar one.
The frequencies of oscillation, computed for specific stellar models of the grid using the {\em Porto Oscillations Code} (\posc), are also provided.

This work was accomplished in order to support the preparation of the CoRoT mission within the {\em Evolution and Seismic Tools Activity} (CoRoT/ESTA). 
On the other hand, the grid can also be used, more generally, to interpret the observational properties of either individual stars or stellar populations. 
The grids (data and documentation) can be found at 
\url{http://www.astro.up.pt/corot/models/cesam}.
\end{abstract}


\section{Introduction}\label{intro}

 The computation of large grids of stellar evolution tracks is particularly useful to study a large number of stars for which photometric and/or spectroscopic observations are available.
 During the last 20 years, both the improvement of the physical description of stellar interiors (in particularly, the opacities and the equation of state) and the computation facilities allowed many groups to produce a huge amount of stellar models.
 These cover from very-low mass stars \citep{Ref1} to higher mass ones \citep{Ref2}, for different values of metal abundances and ages, including the solar standard models \citep{Ref3}.
 It is difficult to imagine that the latest progress in stellar 
population synthesis \citep{Ref4} or on the chemical evolution in the Galaxy 
would have been possible without the help of those grids \citep{Ref5}.
 The application of a large number of stellar evolution models to resolve single stars can, also, be very important in order to constrain the models themselves.
 For example, the comparison between isochrones and Hipparcos Population II stars, in the plan of the HR diagram, pointed out the limitation 
of those models to reproduce the observations of metal poor stars \citep{Ref6}.

 With the advent of helio- and astero-seismology new observational constraints for theoretical stellar models have become available.
 It is very well know how sensitive, for instance, the large and the small frequency separations are to both the global characteristics of stars (mass, radius and chemical composition) and to the theoretical treatment of the stellar internal structure \citep[e.g.][]{monteiro02}. 
 On the other hand, for fixed input physics, the models can be particularly interesting to predict the frequencies of oscillation of stars not (yet!) observed.
 The prediction can be important to select the stellar targets that will be measured.
 If the observational program is space based, those predictions are not only important but absolutely crucial to define the final target list \citep{michel06}.
 The grids reported here have already been used to support the seismic study of pre-main sequence stars \citep{ripepi07,ruoppo07}. 

\begin{table*}[ht]
\caption{Summary of the input physics adopted to calculate the evolution models in the grids described in this work.
 Further details are also provided by \citet{lebreton08}.}
\label{tab:physics}
\centering
\begin{tabular}{lcl}
\hline\noalign{\smallskip}
{\bf Item of the physics} & {\bf Selection} & {\bf Reference(s)} \\[3pt]
\hline\noalign{\smallskip}
Equation of State & OPAL & \citet[2005 Tables]{rogers96} \\
Opacities & OPAL & \citet{iglesias96} + \citet{alexander94} \\
Nuclear reaction rates &	NACRE &	\citet{angulo99} \\
Convection &	 MLT & \citet{vitense58} + \citet{henyey65} \\
Overshoot & {\em none} &	- \\
Diffusion/settling &	{\em none} & 	- \\
Mixture & solar & \citet{grevesse93} \\
Atmosphere & gray &- \\
\noalign{\smallskip}\hline
\end{tabular}
\end{table*}

 In this work we briefly describe the grids of stellar models and their frequencies of oscillations produced in order to support the preparation of the CoRoT Mission \citep{corot06}.
 We start by identifying the physics and parameters that have been used to produce the models and then we discuss the data made available online.

\section{Parameters of the grids}\label{sec:cesam}

 The present grid has been computed using the \cesam\ code \citep{morel97,cesam}, version 2k.
 In order to produce a grid with different initial conditions (birthline) an extension of the above version of \cesam, developed by \cite{marques08}, has been used.

\subsection{The input physics}\label{sec:physics}

 The input physics used to calculate the models is similar to the reference physics adopted by the CoRoT/ESTA group \citep{lebreton08}.
 It is summarised in Table~\ref{tab:physics}.
  The models consider neither microscopic diffusion of chemical elements nor overshoot of the convective core. 

\subsection{Stellar parameters}\label{sec:par}

 We have calculated evolution models with masses ranging from 0.8 to 8 $M_\odot$, from the contraction phase to the beginning of the red giant branch (RGB).
 All models were computed using the solar abundance of metals and helium and also the solar mixing length parameter (see Table~\ref{tab:par} for a summary).
 The zero age main sequence (ZAMS) and the terminal age main sequence (TAMS) in the grids are defined as follows:
 \begin{description}
 \item [\; ZAMS:] when $99\%$ of all energy is produce by nuclear reactions; 
 \item [\; TAMS:] when the abundance of hydrogen at the centre is $X_c = 0.01\pm 0.0001$.
 \end{description}
 
\begin{table}[ht]
\caption{Summary of the stellar parameters used to calculate the evolution models in the grids described in this work.
 Here $M_\odot$ is the mass of the sun.}
\label{tab:par}
\centering
\begin{tabular}{ll}
\hline\noalign{\smallskip}
{\bf Stellar parameter} & {\bf Value}\\[3pt]
\hline\noalign{\smallskip}
Helium abundance & $Y=0.28$ \\
Abundance of metals & $Z/X=0.02857$ \\
MLT parameter & $\alpha=1.8$ \\
Stellar mass & $0.8 \le M/M_\odot \le 8.0$ \\
\noalign{\smallskip}\hline
\end{tabular}
\end{table}

\section{Initial conditions for stellar evolution}\label{sec:3}

 Two options have been considered for the initial conditions of the evolution for each stellar mass.

\begin{figure}[!ht]
\centering
\begin{tabular}{ccc}
  \hskip -0.5cm \includegraphics[width=\hsize,angle=0]{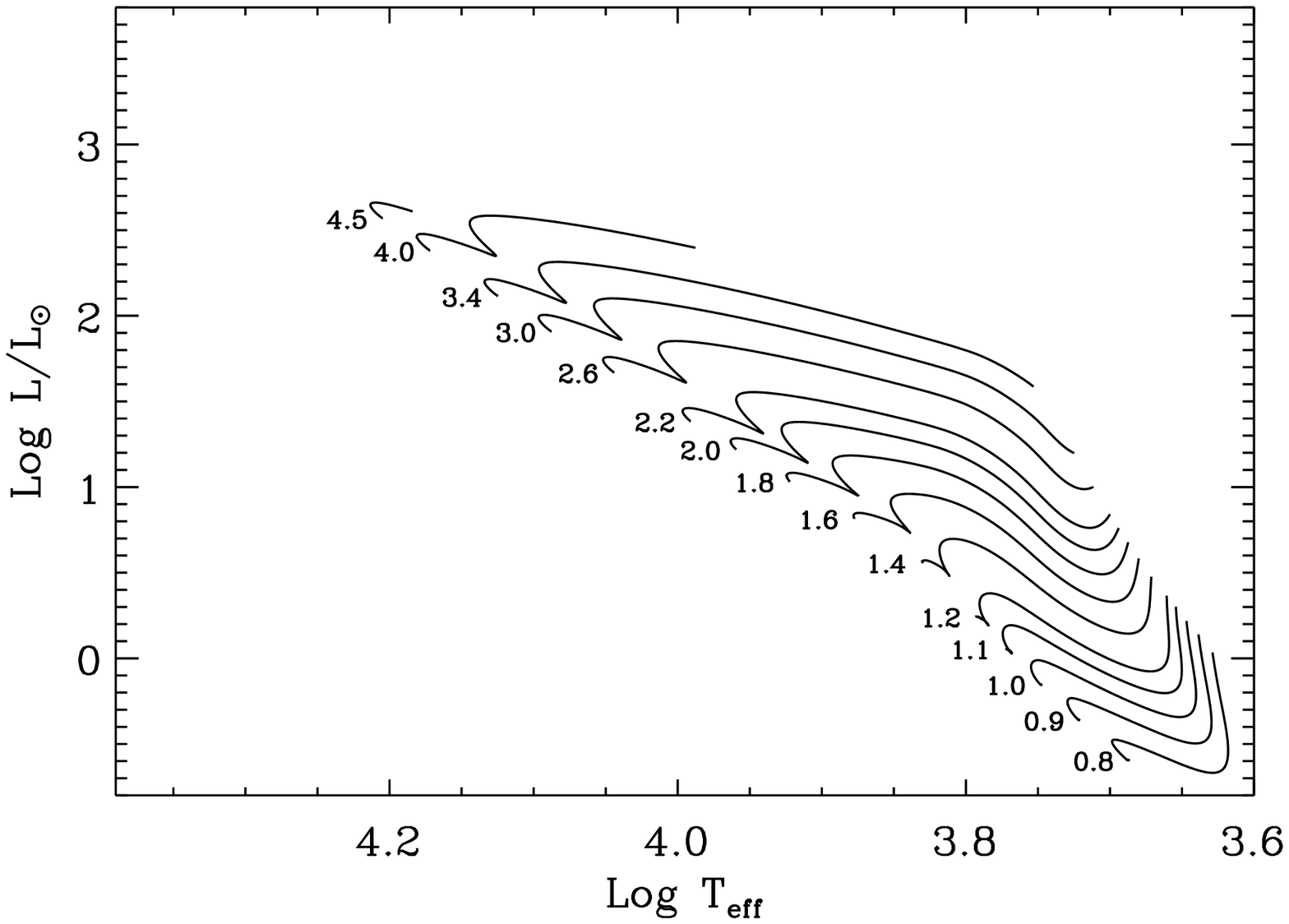} \\
  \hskip -0.5cm \includegraphics[width=\hsize,angle=0]{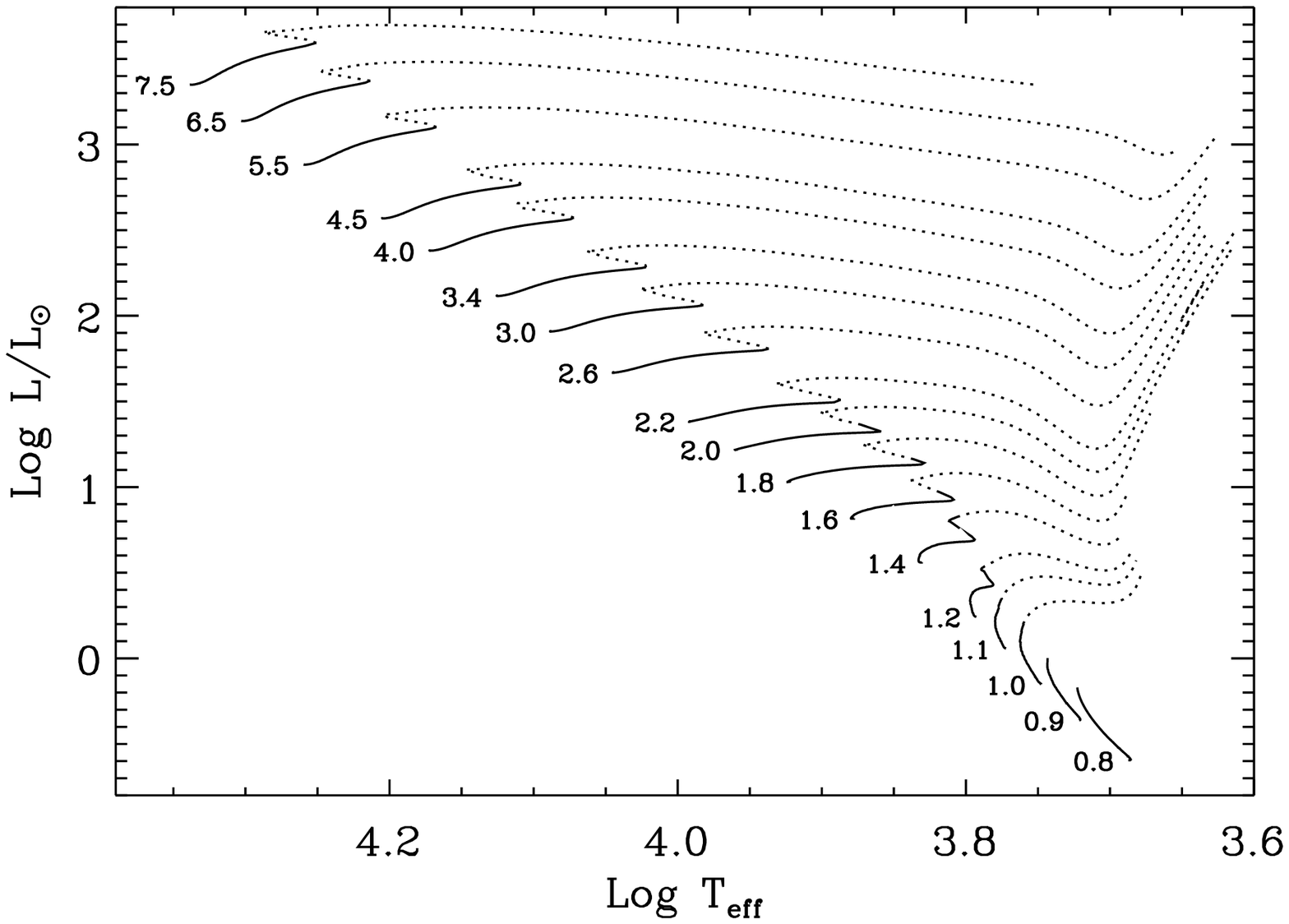}
\end{tabular}
\caption{HR diagrams for models of different masses at different ages: top panel -- pre-main sequence evolution; bottom panel -- main sequence evolution (full lines) and post-main sequence evolution (dotted lines).
Only PMS models with an age above 1~Myr are shown.
The number indicates the mass of the model in solar masses ($M_\odot$).}
\label{fig:grid-a}
\end{figure}

\subsection{GRID A: evolution from fully convective spheres}

 The initial models used for the pre-main sequence (PMS) evolution have tradicionaly been fully convective models with arbitrarily large radii (about $50~R_{\odot}$, e.g. \citealt{1966ARA&A...4..171H,Iben:1965}). 
 The reason behind this is that it was thought that stars were formed by the homologous contraction of the protostellar cloud; this non-hydrostatic phase was assumed to be so fast that the energy radiated was much lower than the gravitational energy of the cloud. 
 A simple energy budget argument \citep{1966ARA&A...4..171H} lead then to this initial large radius.

 This simple picture of star formation is no longer held to be adequate \citep{Palla/Stahler:1991,Palla/Stahler:1992}. 
 Nevertheless, the high convenience of the so-called ``classical'' initial models means that they continue to be widely used; besides, the influence of the initial conditions fades as the star evolves on the PMS.

 The example of the grid that uses this initial condition is illustrated in Fig.~\ref{fig:grid-a}, in particular in the upper panel where the pre-main sequence evolution from an age of about 1~Myr is shown for models of different stellar masses.
 
\subsection{GRID B: evolution from a birthline}

\begin{figure}[!ht]
\centering
\includegraphics[width=\hsize,angle=0]{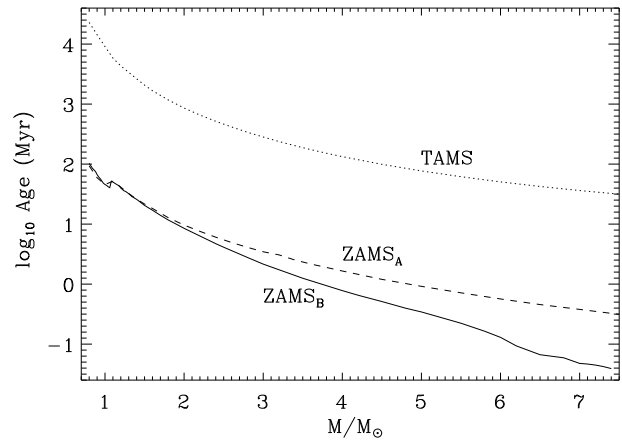}
\caption{
 Stellar age at the ZAMS and TAMS for different stellar masses, in the two grids.
 The age at the ZAMS decreases must faster with stellar mass for the grid using the {\em birthline} (B) than the grid starting from the {\em classical} (politrope) initial condition (A).
 There is no significant difference for the age at the TAMS in the two grids.}
\label{fig:age}
\end{figure}

 It is now known that stars are formed in a very non-homologous way: an hydrostatic core is formed first, on to which matter from the parental cloud accretes, either directly or through an accretion disk.
 Consequently, more realistic initial conditions are used for the calculation of this second grid.
 The most important reason to do so is that the PMS phase becomes shorter with increasing stellar mass, i.e. stars with higher masses are born nearer the ZAMS (see Fig.~\ref{fig:age}). 
 More realistic initial conditions are not relevant for stars with masses lower than $\sim 1.5 M_{\odot}$, but they become more important for stars with higher masses. 
 In particular, stars with masses higher than $\sim 7 M_{\odot}$ are born so near the ZAMS that they do not have a proper PMS phase.
 
 We started the calculation of the initial models with a $0.5 M_{\odot}$ model evolved along the Hayashi track until its radius was the same as the radius of an accreting protostar with the same mass in \cite{Palla/Stahler:1991}. 
 Then accretion is turned on at a constant rate of $\dot{M}=10^{{-}5} M_{\odot}$~yr$^{{-}1}$. 
 The actual mass accretion rates vary with each individual case, and are not in any way constant \citep{2004A&A...419..405S}. 
 In particular, the mass accretion rate seams to increase with increasing mass.
 Given the goal of this grid, it is not necessary to include all the complicated phenomena of stellar birth in the calculation of the initial models; our accretion rate is intended to represent a typical value. 

\begin{figure}[!ht]
\centering
\includegraphics[width=\hsize,angle=0]{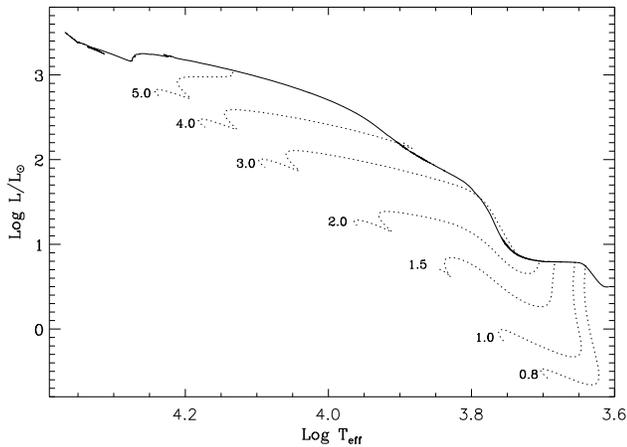}
\caption{HR diagram displaying the birthline (full line) and the pre-main sequence evolution of some models (the number indicates the mass of the model in solar masses $M_\odot$).}
\label{fig:birth}
\end{figure}

 The evolutionary track of the accreting protostar on the HR diagram is called the {\em birthline}.
 An initial model for PMS evolution is obtained when the accreting protostar reaches the mass of the model.
 Both, the birthline and some pre-main sequence evolutionary tracks, are illustrated in Fig.~\ref{fig:birth}.

 An important influence on the location of the birthline is the deuterium content of the parental cloud. 
 The mass-radius relation of the protostar is set to a high degree by deuterium burning.
 In ``classical'' PMS stars, the structural influence of deuterium burning is small: due to its low abundance, the deuterium burning phase is short. 
 In accreting protostars, however, continuous accretion ensures that fresh deuterium is continuously suplied (see \citealt{Palla/Stahler:1991}). 
 Following \cite{2006ApJ...647.1106L}, we select a lower [D/H] number ratio than typically used; this was $\rm{[D/H]}=1.5{\times}10^{{-}5}$, rather than $2.5{\times}10^{{-}5}$ or $3{\times}10^{{-}5}$.
 
\section{Data}\label{sec:4}
 
 The data that has been produced, as part of the grid, includes:
 \begin{itemize}
 \item evolutionary sequences (files of type \url{*.dat} and \url{*.HR}),
 \item stellar models at several ages (files of type \url{*.OSC}),
 \item oscillation frequencies (files of type \url{*.freq}).
 \end{itemize}
 
 All files are available online for download at
 \begin{itemize}
 \item []{\small \url{http://www.astro.up.pt/corot/models/cesam/}}
 \end{itemize}
 
 Specific information or data about the grids can also be provided on request from the authors.
 
\subsection{Evolutionary tracks and stellar models}

\begin{figure}[!h]
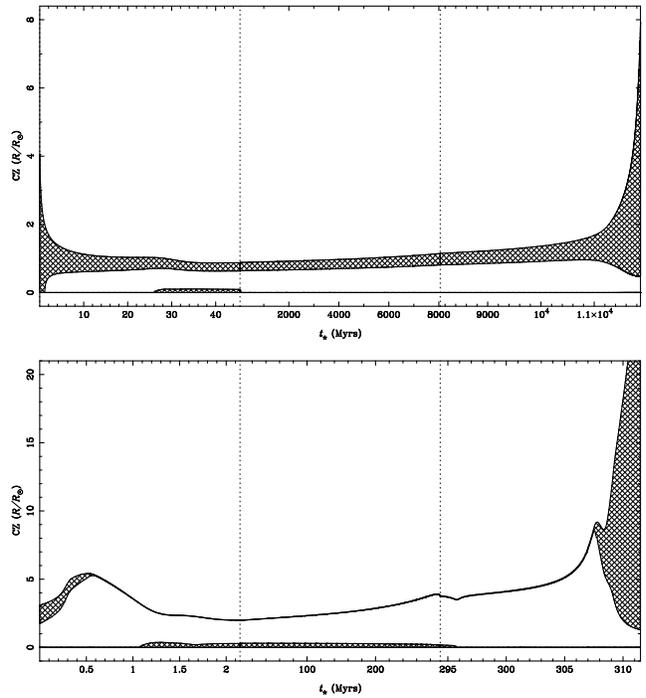

\centering
  \includegraphics[height=\hsize,angle=-90]{zc_10}\\[6pt]
  \includegraphics[height=\hsize,angle=-90]{zc_30}
\caption{Evolution of the size of the convective regions (shaded area) with time, in models of GRID~B, having masses $M{=}1.0~M_\odot$ (top panel) and $M{=}3M_\odot$ (bottom panel).
 The full evolution from the birthline to the RGB are represented (dotted vertical lines represent the ZAMS and the TAMS).}
\label{fig:cz-evol}
\end{figure}

 The evolution of models of different stellar masses are provided using the \url{*.dat} files (in same cases de \url{*.HR} output files from \cesam\ are also included).
 Here the evolution of the global parameters of the star is registered.
 The panels in Fig.~\ref{fig:cz-evol} illustrate, as an example, how the size of the convective zones changes with time for models of $1.0~M_\odot$ and $3.0~M_\odot$ stars.
 The HR diagrams in Figs~\ref{fig:grid-a}-\ref{fig:birth} are also examples of the data available in these files.
 
 For each value of stellar mass, several interior models at different ages can be found in the \url{*.OSC} files.
 For the details on the format and contents of the files please see the documentation available at the website.
 The files with the data on the internal structure of a star (with a specific mass) at a given time are grouped according to the following phases of evolution:
\begin{description}
\item [\; prems] --  pre-main sequence evolution,
\item [\; ms] --  main sequence evolution,
\item [\; postms] --   post-main sequence evolution.
\end{description}
 Each model has a \url{*.OSC} file containing all major quantities describing the internal structure of the star (and used to calculate the seismic properties).

 
\subsection{Frequencies of oscillation}

 For some of the models of the grid the frequencies of oscillation were computed using the \posc\ code \citep{posc}.
 These are the \url{*.freq} files containing the basic mode parameters.
 Only low degree modes are included ($0\le l\le 3$) with mode orders from $n=1$ and with a frequency below the acoustic cutoff frequency of the models.
 In same cases g-modes are also given (identified with a negative mode order).
 
 The boundary condition at the atmosphere used in the computations assumes an isothermal atmosphere at the top.
 For further details please see \cite{posc} and the documentation available at the website.
 
\section{Concluding remarks}

 In this paper we have reviewed the basic aspects of reference grids of stellar models produced with the \cesam\ code \citep{cesam}.
 Two grids are provided with different initial conditions: A - fully convective sphere (politrope); B - the birthline from \citet{Palla/Stahler:1991,Palla/Stahler:1992}.
 The remaining aspects of the physics are fixed as in \citet{lebreton08}.
 The grid also provides frequencies of oscillation for p- and g-modes of some of the models.
 
 The major goal of the grids is to support the preparation and exploitation of space missions for asteroseismology and in particular the French CoRoT space mission \citep{corot06} launched in December 2006.
 All data is avaliable online\footnote{\url{http://www.astro.up.pt/corot/models/cesam/}}.

\acknowledgements

 This work was supported in part by the European Helio- and Asteroseismology
Network (HELAS), a major international collaboration funded by the
European Commission's Sixth Framework Programme.
 JPM and MJPFGM acknowledge the support by FCT and FEDER (POCI2010)
through projects {\small POCI/CTE-AST/57610/2004}, {\small POCI/V.5/B0094/2005} and {\small PTDC/CTE-AST/65971/2006}.


\end{document}